\documentstyle[12pt]{article}
\begin{document}
\baselineskip .3in
\begin{titlepage}
\begin{center}{\large{\bf A Study on the Structure of the Proton  }}

\vskip .2in

\end{center}
\vskip .1in

\vskip .3in

The structure function of the proton has been investigated and has been found to possess the power law behaviour in conformity with the empirical fits to the experimental findings. We have estimated F$_{2}$(x, Q$^{2}$)/F$_{2}$(x, Q$_{0}$$^{2}$) with the anomalous dimension D$_{A}$ predicted from the statistical model as an input and the result is found to be in good agreement with the recent data available in the deep inelastic region.
\vskip .1in

\end{titlepage}

It is wellknown that scaling laws, power laws in particular, have fundamental importance as they often reflect a deep symmetry in the underlying Physics and scaling laws with anomalous exponents arise in systems that are described by fractals. The fractal characteristics of a hadron were studied and analysed by the authors in the context of the statistical model of a hadron [1,2].In the context of the model the fractal dimension D$_{F}$ of a hadron in terms of the anomalous dimension $D_{A}$ and the topological dimension D$_{T}$ where D$_{A}$ = D$_{F}$ - D$_{T}$, have been suggested [1,2]. It is relevant to recall in this connection that Wilson's many scales of length approach was designed to deal with the explicit scale dependence and the similarity between fractals and the renormalization group (RG) has been emphasized by Mandelbrot[3]. The RG formalism provides the necessary framework to deal with the fractal properties. Such mutiple scales of length appear when there is a cross-over from weak coupling at short distances to strong coupling at nuclear distances and as such the RG in the form of the Callan-Symanzik invariant of the Gell-Mann Low equation has played a prominent role in the development of QCD. It is used to show that the non-Abelian gauge theories are asymptotically free which means that the short-distance couplings are weak but increase with the length scale. The RG framework can explain qualitatively the weak coupling evident in the analysis of the results of deep inelastic scattering (dis)of electrons-protons and the strong coupling evident in the binding of quarks to form proton.

The Statistical model suggested for a hadron is based on some assumptions:
A hadron is assumed to be consisting of a sea of virtual $q \bar q$ pairs in addition to the valence partners which  determines the quantum number of the colourless hadron.
 The quarks real and virtual are assumed to be of same flavour and colour so that they may be regarded as identical and indistinguihable i.e.the real and virtual quarks are treated in the same footing.The indistinguishability of the valence quark with its corresponding virtual partners calls for the existence of the quantum-mechanical uncertainty in its available phase space. 
The valence quarks(antiquarks) are regarded as (approximately) non-interacting and moving independently and without any correlation in conformity with the experimental findings. However in the framework of the statistical model the valence quark is considered to move in an average smooth background potential due to its interaction with the virtual partners in the sea.
With the above consideration we come across an expression for the probaility density of the proton as:
\begin{equation}
 {\mid \psi(r) \mid}^{2} = A{(r_0 -r)}^{\frac{3}{2}} \Theta(r_0 -r)
\end{equation}
where $r_{0}$ is a parameter characterzing the size of a hadron and $A = \frac{\rm 315}{\rm 64 \pi r_{0}^{\frac{\rm 9}{\rm 2}}}$. With $\xi$= r$_{0}$-r we can recast the expression (1) as:
\begin{equation}
{\mid \psi(r) \mid}^{2}\equiv P(\xi) \sim {\xi}^{3/2}\Theta(\xi)
\end{equation}
The above expression can be expresed as;
\begin{equation}
P(\xi)\sim {\xi}^{3/2}\Theta(\xi)\sim  {\xi}^{D_{F}-D_{T}}\Theta(\xi)\sim{\xi}^{D_{A}}\Theta(\xi)
\end{equation}
Where $D_{T}$ is the topological dimension = 3, $D_{F}$ is fractal dimension= 9/2 and $D_{A}$ = 3/2 is the anomalous dimension. The aforesaid function is not a single valued function and is thus non-analytic as its derivative does not exist. There is a branch cut with a branch point singularty at $\xi$ = 0. Thus it is both non-differentiable and scale dependent. Non differentiability implies scale divergence. The power law function $P({\xi})$ is a scaling function as it satisfies the homogeneity relation P(${\Lambda \xi}$)= $\Lambda^{D_{A}}$$P({\xi})$ for all positive values of the scale factor $\Lambda$. According to the idea of fractal we have suggested that proton has a fractal dimension 9/2 a the boundary or in other words boundary of the proton possesses fractal structure having fractal dimension $D_{F}$=9/2 and a hadron is described as a self similar fractal object of $D_{F}$=9/2.The anomalous dimension   $D_{A}$=$D_{F}$-$D_{T}$ is obtained as = 3/2 for a proton.

{{\large \bf The Structure Function:}}

In the deep inelastic region the cross-section for process ep$\rightarrow$eX the symmetric hadronic tensor $W_{{\mu}{\nu}}(p,q)$ can be expressed  in terms of structure function W$_{1}$ and W$_{2}$ as:
$W_{{\mu}{\nu}} = (-g_{{\mu}{\nu}}$
$+ q_{\mu}q_{\nu}/q^{2})$ 
$W_{1}(\nu, q^{2}) + $
\vskip .1in
${\frac{\rm 1}{\rm M^{2}}}$
$(p_{\mu} - {\frac{\rm pq}{\rm q^{2}}}$
$q_{\mu})(p_{\nu} - {\frac{\rm pq}{\rm q^{2}}}$
$q_{\nu})W_{2}(\nu, q^{2})$

\vskip .1 in
where \begin{equation}
\ q^2 =(k-k')^2=-Q^2 and  Q^2>0
\end{equation}
and s = $(p+k)^2$, $\nu$ = p.q/M, W$^{2}$ = p$^{2}_{x}$, M is the proton mass, $W_{1}$($\nu$,q$^{2}$), $W_{2}$($\nu$,q$^{2}$) are called structure functions.
The deep inelastic zone is defined as $\nu$$>>$M,  -q$^{2}$ $>>$ M$^{2}$, $x = -q^2/{2\nu.M}$ = $Q^2/{2\nu.M}$ and they are finite. The F$_{2}$ structure function is given by F$_{2}$ = $\nu$ $W_{2}$($\nu$,Q$^{2}$)
  For fixed value of x, the F${_2}$- structure function is independent of momentum transfer q$^{2}$ which is referred as exact scaling. But the experimental results in the deep inelastic zone show scaling violation. It would be interesting to study the variation of F$_{2}$ (x, Q$^2$) with respect to $-q^2/2q.p$ or, x in the dis domain through the RG equation.We know that the RG equation has been put in a particularly simple form by Callan and Symanzik (CS). To investigate the scaling behaviour of the F$_{2}$ (x, Q$^2$)we assume that F$_{2}$ (x, Q$^2$) varies with the scale in such a way that $F_2$ itself is the only relevant parameter on which the physical laws depend whatever the scale. This means that the variation of $F_2$ with scale will depend on $F_2$ alone through some function, known as the CS $ \beta$ function. Hence we may use the CS type of the RG equation as: 
\begin{equation}
\frac{\rm dF_{2}(x, Q^{2})}{\rm d[ln x]} = - \lambda F_{2}(x, Q^{2})
\end{equation}
where $\lambda$ is a constant with respect to x and the characteristic scale for the variation of x is unity. Integrating we get;
\begin{equation}
F_{2}(x, Q^{2}) \propto x^{-\lambda (Q^{2})}
\end{equation}
\begin{equation}
or,\frac{\rm dF_{2}}{\rm dx} \propto F_{2}/x
\end{equation}
 
Hence we find that $\bf \frac {\rm \delta[ln F_2]}{\rm \delta[ln x]}$ is independent of x. In H1 collaboration, Adloff et al [4] have recently reported that for the x dependence of $F_2$ at low x(for$ x\leq {.01}$)is consistent with a power law behaviour as $F_2  {\alpha} x^{-\lambda}$ for fixed $Q^2$ and that the rise of $F_2$ i.e. $\frac{\rm \delta F_2}{\rm \delta x}$ is proportional to $F_2/x$ in conformity with the relations derived in (6) and (7) respectively. It may be mentioned that below the dis region for fixed $Q ^2<1GeV^2$, the Regge phenomenology predicts $F_2(x,q^2)\propto x^{-\lambda}$ where $\lambda = \alpha_{\rho}(0)-1\equiv .08$ is given by the Pomeron intercept and is independent of x and $Q^2$. Adloff et al [4] have also been found $\delta [ln F_2]/\delta [ln x]$ as a function of both x and $Q^2$ and have observed it to be independent of Bjorken x for $x \leq {.01}$ and $Q^2$ between 1.5 and 150 $GeV^2$.
Similarly considering (Q$^{2}$ /$Q_{0}^{2}$) as the scale we may write the RG equation for the F$_{2}$ structure function through the $\beta$ function as in Nottale[7] as:
\begin{equation}
\frac{\rm dF_{2}(x, Q^{2})}{\rm dln(Q^{2}/Q_{0}^{2})} = - a F_{2}(x, Q^{2})
\end{equation}
where Q$_{0}^{2}$ corresponds to the square of the momentum transfer for probing the proton at the valence level and Q$^{2}$ is some other value of the square of the momentum transfer such that Q$^{2}$ $>$ Q$_{0}^{2}$, 'a' is a constant and $Q_{0}^{2}$ is the characteristic scale. Integrating the above expression we get;
\begin{equation}
F_{2}(x, Q^{2}) =   A ({\frac {\rm Q^{2}}{\rm Q_{0}^{2}}})^{-a}
\end{equation}
where A is an integration constant. Hence we get,
\begin{equation}
F_2(x,Q^{2})=F_2(x,{Q_0}^{2})({Q_0}^2/Q^2)^a
\end{equation}
with
\begin{equation}
R = F_2(x, Q^{2})/F_{2}(x,Q_{0}^2) 
\end{equation}
We have;
\begin{equation}
R=(Q_{0}^2/Q^2)^{a}
\end{equation}
 Hence we observed that 'a' appears as the exponent in the structure functions which is responsible for the violation of the scaling behaviour and it is termed as the anomalous dimension when 'a' is constant [7]. However, to be more general, we may assume the dependence of 'a' on x through a series expansion for low value of x such that a $\simeq$ D$_{A}$-$\gamma$x, where $\gamma$ is a constant and we can recast the above expression as;
\begin{equation}
R=(Q_{0}^2/Q^2)^{D_{A}-\gamma x}
\end{equation}
 The recent experimental data by Adloff et.al. [9] suggested exact scaling for large $Q^2$ and very low x values (corresponding to the Bjorken scale symmetry), so that for x = .0398, F$_{2}$ has the same value = .506 for  Q$^2$ = 90 GeV$^2$ and $Q^2$ = 60 GeV$^2$ respectively.which gives R=1. Thus we have from the expression (13) $\gamma$-D$_{A}$/x = 0. With D$_{A}$ =3/2 i,e, the anomalous dimension for the proton suggested from the statistical model as an input we have $\gamma$ = 1.5/.0398. With the input of this value of $\gamma$ in the expression (13) for another set of values for x = .032; Q$^{2}$ = 150GeV$^2$ and ${Q_{0}}^2$ = 120 GeV$^2$, we get R$\equiv$ F$_{2}$(.032, 150)/F$_{2}$(.032, 120) $\approx$ .938 which agrees reasonably well with the  corresponding experimental value of the ratio $R=.550/.558=.985$ as our error is estimated to be  $<5\%$. Perkins et al[8] have observed that the neutrino and electromagnetic data show similar patterns of the scaling violation and it may have power law rather than logarithmic behaviour. Again we find from (13) that $\frac{\rm \delta ln {F_{2}(x,Q^2)}}{\rm \delta ln x}\mid _{Q^2} =  -\gamma x ln ({Q^2/{Q_0}^2})$ indicating that the logarithmic slope rises with $ ln {Q^2}$. Similar behaviour has been observed by Adloff et al [4] in H1 collaboration for low x domain of dis of positron proton system for $5\star 10^{-5}\leq x\leq .01$ and $Q^2\geq 1.5 GeV^2$. They have also found that the aforesaid logarithmic derivative is proportional to x as observed by us. Moreover, the QCD extrapolations and fits for $x>10^{-2}$ also suggest that it rises with x.

In the present work we have seen that the two power-law behaviour of $F_{2}(x,Q^2)$ of the proton can be derived as the solution of the CS type RG equation and they are similar to the emperical behaviour of $F_{2}(x,Q^2)$ suggested from the experimental data as discussed. This may be an indication of the fractal structure of the proton. Moreover, the assignment of anomalous dimension D$_{A}$ = 3/2 from the statistical model for  $F_{2}(x,Q^2)$ leads to a reasonable agreement of our theoretical estimate of the ratio R with the corresponding experimental findings suggesting  that the scaling violation  of the structure function of the proton may be attributed to the anomalous dimension of the proton and that proton may be described as a scaling fractal with fractal dimention 9/2.

{\bf References}:-

\noindent
[1]S.N.Banerjee et al,Int.J.Mod.Phys.A16(2001)201.
\noindent
[2]S.N.Banerjee et al,Int.J.Mod.Phys. A17(2002)4939,Can.J.Phys.80(2002)787,
Bhattacharya et al., Int.J. Mod.Phys.A 15(2000)2053.
\noindent
[3]B.Mandelbrot,The Fractal Geometry of Nature(Freeman,San Francisco,1982).
\noindent
[4]C.Adloff et al,Phys.Lett.B 520(2001)183.
\noindent
[5]S.N.Banerjee et al,Ann. Phys.(N.Y.).150(1983)150.
\noindent
[6]F.E.Close,Anm Introduction to Quarks and partons(Academic Press,London,1979).
\noindent
[7]L.Nottale,Fractal Space-time and Microphysics(World Scientific,Singapore,1993).
\noindent
[8]D.H.Perkinsetal,Phys.Lett.67B(1997)347.
\noindent
[9]C.Adloffetal,Eur.Phys.Journ.C21(2001)33.

\end{document}